# Numerical Investigation of Mechanical Properties of Aluminum-Copper Alloys at Nanoscale

Satyajit Mojumder[1,2], Md Shajedul Hoque Thakur[2], Mahmudul Islam[2], Monon Mahboob[2], Mohammad Motalab[2*]

[1]Theoretical and Applied Mechanics Program, Northwestern University, Evanston, IL-60208, USA

[2]Department of Mechanical Engineering, Bangladesh University of Engineering and Technology, Dhaka-1000, Bangladesh.

**Abstract**

Nanoindentation is a powerful tool capable of providing fundamental insights of material elastic and plastic response at the nanoscale. Alloys at nanoscale are particularly interesting as the local heterogeneity and deformation mechanism revealed by atomistic study offers a better way to understand hardening mechanism to build a stronger material. In this work, nanoindentation in *Al-Cu* alloys are studied using atomistic simulations to investigate the effects of loading direction, alloying percentages of *Cu* via dislocation-driven mechanisms. Also, a low-fidelity finite element (FE) model has been developed for nanoindentation simulations where nanoscale materials properties are used from atomistic simulations. Material properties, such as hardness and reduced modulus, are computed from both the FE and MD simulations and then compared. Considering the fundamental difference between these two numerical approaches, the FE results obtained from the present study conform fairly with those from MD simulations. This paves a way into finding material properties of alloys with reduced simulation time and cost by using FE where high-fidelity results are not required. The results have been presented as load-displacement analysis, dislocation density, dislocation loops nucleation and propagation, von-Mises stress distribution and surface imprints. The techniques adopted in this paper to incorporate atomistic data into FE simulations can be further extended for finding other mechanical and fracture properties for complex alloy materials.

*Corresponding author *Email address:* abdulmotalab@me.buet.ac.bd, Phone: +8801779198595

# 1. Introduction

Aluminum is one of the major engineering materials and has wide varieties of application in modern technology.[1–3] Although pure *Al* is a good conductor for electricity and heat, it is a very soft material which restricts its application in engineering fields that demand high mechanical strength. On the other hand, alloying the *Al* with different solute atoms such as *Cu, Mg, Zn, Si, Mn, Sc* can improve the properties significantly and makes it possible to apply in different applications such as automobile industry,[4] naval engineering,[5] cryogenics,[6] welding technology[7] and additive manufacturing.[8] *Al-Cu* alloy is one of the major alloys of *Al*. Cu is the primary alloying element of 2000 series Aluminum alloy. Once heat treated, this alloy shows similar mechanical properties to that of mild steel and significant corrosion resistance.[9] Researchers have implemented molecular dynamics simulations to study *Al-Cu* alloy in recent years.[10,11] The addition of *Cu* enhances the material properties by the solid solution strengthening and strain hardening. Again, Ma *et al.*[12] studied lattice misfit due to the *Cu* solute atom in *Al* metal and concluded that strengthening capability is highly dependent upon the solubility. The inclusion of *Cu* in *Al* structure is found to show improved corrosion resistance and high strength.[13] This alloy is generally used for construction purposes such as vehicle bodies,[14] ships,[15] pressure vessels, cylindrical tanks etc. Applications of this alloy have found its way to nano and micro-electronics, which provide further motivation for detailed study of this alloy in nanoscale.[16]

Nanoindentation, also known as instrumented indentation, has emerged as a powerful tool for the measurement of localized mechanical properties of materials at micro and nanoscale. Nanoindentation also provides useful insights about the shear instability, dislocation source activation, dislocation propagation, phase transformation along with the fundamental material properties such as elastic modulus and hardness.[17,18] In recent years, nanoindentations are performed on bone, tissue etc. to measure their stiffness and other mechanical properties for application in biomedical science.[19] It can also be used to measure the local property of both homogeneous[20] and heterogeneous materials.[21,22] Furthermore, nanoindentation can provide in depth understanding of incipient plasticity and materials yielding through the dislocation nucleation and propagation in different nanomaterials. Dislocations in *Al* are pinned through interactions with *Mg* atoms and thus higher stresses is required for dislocation movement.[23] Hence the dislocation propagation during nanoindentation in *Al-Mg* alloy should also depend on the distribution of the *Mg* atom. Similar effects should also be applicable for *Al-Cu* alloys. Detailed study of *Al* alloys such as *Al-Cu* nanoindentation is therefore critical to understand the mechanism of dislocation propagation in *Al* alloy.

Molecular dynamics (MD) simulation can be an effective method to study nanoindentation in order to understand different mechanical properties of nanomaterials. The applicability of atomistic approach of nanoindentation to measure material properties and identify incipient plasticity were investigated by Landman *et al.*[24] MD simulations of Si and SiC nanoindentation showed that, nanoindentation can trigger phase transformation during the loading process.[25–27] This phase change due to the loading of nano-indenter has been found to have strong dependency with the temperature.[25] With the advancement of computational capability, researchers implement MD simulations to explain and understand experimental results of nanoindentation.[28] Dislocation pattern in *Al* <111> surface for different interatomic potentials has been investigated by Lee *et al.*[29]. They explained the nucleation sites, dislocation locks and loops formation just underneath the indenter tip and the prismatic dislocation loops far away from the contact surface. MD study on *Al* nanoindentation found that, higher temperature results in pre-nucleation of dislocations.[30] MD study of nanoindentation of pure *Al* revealed the effects of indenter speed, depth and size on dislocation nucleation and propagation.[31] It is concluded from the study, that the surface roughness of the indenter can affect the nature of dislocation nucleation. In case of nanoindentation of *Mg*, the experimental and MD approach both reveal that, indentations on the basal plane have higher pop-in load and higher displacement than in prismatic plane.[32–34] Recent works on nanoindentation of polyethylene are focused on calculating their hardness through MD simulation.[35] So, these extensive nanoindentation studies through MD simulations point out to the fact that, atomistic approach of nanoindentation for *Al* alloys is expected to yield accurate and consistent material properties at nanoscale.

However, one of the major limitations of MD simulation is that it is computationally expensive compared to other numerical approaches of nanoindentation. On the other hand, Finite Element Method (FEM) is computationally less demanding while it can predict the continuum scale of nanoindentation reasonably well. In recent years, there have been a wide implementation of FEM for nanoindentation simulations to characterize the mechanical properties at the bulk scale.[36] Modeling nanoindentation for bulk material and thin film using FEM has been conducted by Bressan *et* al.[37] Their modeling was conducted by using axisymmetric CAX4R element with a cylindrical substrate. FEM studies showed that, the modulus and yield strength of material significantly influence the load-displacement curve (known as P-h curve).[38] Using the FEM technique, it is possible to model the nanoindentation for the nanoscale material although if the length scale is below 10 nm, continuum approach of FEM is no longer useful. This limitation has been previously countered by appropriate scaling methods.[39] However, the major challenge for FEM is the availability of experimental tensile test data for the material at the nanoscale. Though the tensile test data of the materials are more or less known for common engineering materials, their mechanical properties such as yield stress, fracture strain, etc. have different values at the nanoscale[40–42] which can be attributed to the size effects. Therefore, modeling of nanoindentation using FEM, where the indentation depth is as low as few nanometers, will give erroneous results and cannot be compared with the atomistic results. To solve this problem, the mechanical properties obtained from MD tensile test can be used in FEM approach of nanoindentation.[43] Using this same approach, Mojumder *et al.*[31] conducted a systematic study of pure *Al* nanoindentation for different crystallographic orientations, indentation depths, indentation speeds and indenter sizes. Previously, a similar approach was implemented by Vodenitcharova *et al*.[43] in order to validate their nanoindentation results of FEM for the few nm indentation in *Al*, which found a good agreement between the *P-h* diagrams of MD and FEM simulations. Their study provides a pathway to obtain the materials properties for nanoscale materials without performing large scale atomistic simulation using FEM in conjunction with MD. However, this approach has not yet been implemented on metal alloys to investigate the mechanical properties and the effects of alloying percentage on the dislocation behavior during nanoindentation.

In the present study, a systematic investigation of the atomistic nanoindentation of *Al-Cu* alloy for different crystallographic orientations and alloying percentages of *Cu* has been conducted. The modeling method of *Al-Cu* alloy for MD study has been adopted from numerous previous works.[10,11,44,45] The effects of these parameters on the dislocation nucleation, propagation mechanism and material properties such as hardness and reduced modulus have been discussed (section 3.1). Then, tensile test simulations have been performed to obtain the mechanical properties such as elastic modulus, yield strength and Poisson's ratio of the alloy material (section 3.2). Using these material properties, FEM nanoindentation simulations have been carried out to compare the FEM results with that obtained from MD simulations (section 4).

## 2. Methodology

In this study, we have employed both MD and FE simulations to understand the effects of different crystallographic orientations and alloying percentages on the nanoindentation of *Al-Cu* alloy. In order to compare the atomistic results with the FE simulations we have used the material properties obtained from MD tensile tests as the input parameters of the FE simulations.

### 2.1 Atomistic simulation procedure

For the atomistic simulation, a box of FCC *Al* atoms with appropriate dimensions is created. Then the *Al* box has been divided into a number of chunks (of 0.2023 nm thickness each) along *z*-direction. Finally, the *Al* atoms in each chunk are randomly replaced by *Cu* atoms based on their weight percentages, which is varied in the range of 0 - 10%. All the alloy structures modelled in the present study have been generated using LAMMPS Input Structure Generator for Functionally Graded Material (FGM)[46] tool in nanoHUB. In the present study, we consider three different crystallographic orientations as for loading in <0 0 1> direction (<1 0 0> <0 1 0> <0 0 1>) , <1 1 0> direction (<1 1 1> <1 1 $\bar{2}$> <$\bar{1}$ 1 0>), and <1 1 1> direction (<$\bar{1}$ 1 0> <$\bar{1}$ $\bar{1}$ 2> <1 1 1>) in our present simulations. The dimensions and total number of atoms of the simulation box used for these three different crystallographic orientations are presented in Table 1.

Table 1: Simulation parameters used for the atomistic simulations

| Crystallographic orientation | *Cu* % | Simulation box dimension | Number of atoms |
|---|---|---|---|
| <001> | 0,1,2,5,10 | 19.74 nm ×19.74 nm ×12.96 nm | 307328 |
| <110> | 0,1,2,5,10 | 19.41 nm ×19.76 nm ×12.74 nm | 302400 |
| <111> | 0,1,2,5,10 | 19.62 nm × 19.76 nm ×12.39 nm | 298080 |

As shown in Fig. 1, the simulation box is divided into two distinct regions. The spherical indenter penetrates the upper region. The bottom region of 2 nm thickness provides a rigid support for the substrate. Also, this region consisting of fixed atoms (also called Newtonian atoms) functions as a heat bath during the penetration of the indenter in the upper region. The rigid indenter (virtual indenter in LAMMPS[47]) of 5 nm radius is then set up over the substrate and the indenter was pushed into the material, as shown in Fig. 1. The indenter exerts a force of magnitude, $F(r) = -K(r-R)^2$ on each atom where $K$ is the specified force constant, $r$ is the distance from the atom to the center of the indenter, and $R$ is the radius of the indenter. The force is repulsive and $F(r) = 0$ for $r > R$. In all of the simulations force constant is considered as 1 eV/Å$^3$. The loading step is followed by an unloading step adopting displacement control of the indenter. After every time step, the system has been minimized using conjugate gradient method in order to maintain the quasi-static loading process at 0 K temperature. The speed of the indentation is set to 10 ms$^{-1}$ and the indentation depth is kept as 2 nm. Previous study conducted by Mojumder *et al.*[31] showed that these values of indenter speed and indentation depth are reasonable choices.

The interactions between all the atoms within the simulation domain are described by the embedded atom method (EAM) potential,[48,49] which was used extensively for the *Al-Cu* alloy previously.[50–53] In this method, the potential energy of an atom, *i*, is given by:

$$E_i = F_\alpha \left( \sum_{i \neq j} \rho_\beta(r_{ij}) \right) + \frac{1}{2} \sum_{i \neq j} \varphi_{\alpha\beta}(r_{ij}) \tag{1}$$

where, $r_{ij}$ is the distance between atoms *i* and *j*, $\varphi_{\alpha\beta}$ is a pairwise potential function, $\rho_\beta$ is a functional specific to the atomic types of both atoms *i* and *j*, so that different elements can contribute differently to the total electron density at an atomic site depending on the identity of the element at that atomic site, and $F_\alpha$ is an embedding function that represents the energy required to place atom *i* into the electron cloud. $\alpha$ and $\beta$ are the element types of atoms *i* and *j* respectively.

The present nanoindentation procedure have been previously verified by comparing the MD load-displacement curves in case of pure *Al* with Hertzian contact theory for all three directions considered in the present study.[31] The hardness of the material, *H*, is defined as below[54]:

$$H = \frac{P}{S} \tag{2}$$

The projected contact area is calculated with equation 3[55,56] in plastic deformation region:

$$S = \pi(2R - h)h \quad (3)$$

where *h* is the instantaneous depth at which the hardness is being calculated and *R* is the indenter radius. The mean value of *H* over the plastic deformation region is reported in this work. The reduced modulus has been calculated by fitting the elastic deformation region of the load-displacement curve to the power law of the Hertz theory.[56] This method of calculating hardness and reduced modulus had also been implemented by Fu *et al.*[57]

As the indenter penetrates the materials, plastic deformation occurs, and dislocation loops are formed. We have calculated dislocation density using OVITO[58] as a measure of plastic deformation in the indentation process using the DXA algorithm. The dislocation segments have been identified using a trial circuit of 14 x 9 and total dislocation length is then divided by the total volume of the substrate to obtain the dislocation density.

In order to simulate the nanoindentation problem in FE platform, the material properties obtained from the MD tensile test simulations have been used as input properties. The uniaxial tensile test simulations are performed for *Al-Cu* nanowires with the orientations mentioned above and different *Cu* weight percentages. The nanowires have circular cross section and a diameter of 5 nm and a length of 50 nm along *z*-direction, respectively. The loading direction is kept similar to the nanoindentation simulation. The aspect ratio (height: width) of all the nanowires is kept constant as 10:1 and the tensile load is applied in the *z*-axis of the co-ordinate system (crystal directions of <0 0 1>, <1 1 0> and <1 1 1>). First, the initial geometries of the nanowires are created and the pressure of the system is equilibrated by applying the isothermal-isobaric (NPT) ensemble in *z*-direction at 1 bar and a temperature of 300 K for 100 ps. Finally, a uniaxial strain is applied along the *z*-direction of the nanowire at a constant strain rate of $10^8$ s$^{-1}$. The timestep chosen for all the simulations is 1 fs. From the tensile simulations, the stress-strain curve is obtained and the elastic modulus, yield stress and Poisson's ratio are calculated. These results of tensile tests are then further used as the input properties in the finite element (FE) simulations. All the MD simulations are performed using LAMMPS,[47] and visualization is done using OVITO.[58]

## 2.2 FE simulation procedure

The purpose of the FE simulation is to get a reasonable estimation for the materials' properties such as hardness and reduced modulus with a lower computation time than atomistic simulation. Since in the FE simulations, the *Al* and *Al-Cu* alloys are modeled as isotropic homogeneous materials, this simplified the simulations and reduced the 3D problem of nanoindentation to an axisymmetric problem. ABAQUS/Standard[59] has been used for the FE simulations considering a deformable axisymmetric material model with a rigid indenter. The dimensions (20 nm radius, 20 nm height) of the substrate considered is the same as the atomistic simulation model and the input materials properties (elastic modulus, Poisson's ratio, yield stress) are obtained from the molecular dynamics tensile tests of *Al-Cu* alloy nanowire as described in section 2.1. The material is modeled using the four-node axisymmetric element with reduced integration (CAX4R) and the indenter has been considered as analytically rigid. The indenter is pushed into the materials' substrate using a displacement control boundary condition. The left vertical edge of the materials has been considered as the symmetric edge and the bottom of the substrate is fixed as shown in Fig. 3. The top and right vertical edges are kept free to resemble the free surface. We choose 13299 elements as an independent grid and performed all finite element simulations for that. A fine mesh is used in the contact region of the substrate and indenter and a coarse mesh far from the indentation zone. The mesh and numerical code used for the present study has been validated previously by Mojumder *et al.*[31] From the load-displacement curve obtained from the FE simulation, the similar procedure has been used as atomistic calculation for the calculation of the materials properties such as hardness and reduced modulus.

## 3. Results and Discussion

### 3.1 Atomistic results of nanoindentation on *Al-Cu* alloys

#### 3.1.1 Effects of alloying on P-h curves and dislocation density

The load–displacement curves for different alloying percentages of *Cu* in *Al* are shown for different loading directions considered in the present study in Fig. 4. In case of all considered loading directions, it is observed that the load carrying capacity can both increase and decrease depending on the inclusion percentage of *Cu* in the alloy. For <0 0 1> loading direction, the load carrying capacity increases for 1% *Cu* percentage compared to pure *Al*. Beyond 1% *Cu* percentage, the load carrying capacity gradually decreases with increasing *Cu* percentages. In case of <1 1 0> loading direction (Fig. 4(b)), the load carrying capacity remains somewhat similar for different percentages of *Cu* inclusion. When the loading direction is <1 1 1>, we can observe increase in load carrying capacity for 1% and 2% *Cu* inclusion. This is due to the fact that, as the foreign atoms of *Cu* replaces the host atom of *Al* in the alloy, it can result in solid solution hardening or softening depending on the interaction between *Al* and *Cu* atoms. The atomic misfit of *Al* and *Cu* crystals are responsible for this. At 0 K, the lattice constants of *Al* and *Cu* are 4.05 Å and 3.61 Å, respectively. Hence the lattice misfit between *Al* and *Cu* is ~11%, which significantly alters the material properties of *Al-Cu* alloy from pure *Al*. This phenomenon of material property alteration due to lattice misfit has been reported in previous studies.[60] Furthermore, the atomic size of *Cu* is larger than the atomic size of the *Al*. Previous analysis[56] showed that, a higher load carrying capacity can be achieved if the number of defect and dislocation nucleation events is kept at minimum during plastic deformation. Hence, a proper look at the dislocation density variation during nanoindentation is necessary to understand the hardening/softening mechanism. In Fig. 5, the dislocation density for different percentages of *Cu* addition is shown for different orientations of loading. It can be seen that, for <0 0 1> loading direction, pure *Al* results in minimum dislocation density during nanoindentation. But for <1 1 0> direction, 1% and 2% *Cu* inclusion result in lesser dislocation density compared to pure *Al*. The same is observed for <1 1 1> loading direction. It is also observed from Fig. 5 (a), (b) and (c) that, when the loading direction is <1 1 0> the dislocation densities for different *Cu* alloy percentages are much lower than those of the other two loading directions considered in the present study. From Fig. 5, it is evident that, there are no dislocations during the initial indentation periods, in case of all loading direction as these periods implies the initial elastic deformation of Fig. 4. When the displacement of the indenter reaches from 0.25 to 0.75 nm, the dislocation density starts to increase with the displacement following somewhat a linear trend. This trend continues throughout the nanoindentation process in case of <1 1 0> loading direction. So, the loading direction and *Cu* inclusion percentage have significant implications on the P-h curve and dislocation density.

### 3.1.2 Effects of alloying on formation of dislocation loops

In Fig. 6, the dislocation loops are shown for the different percentages of *Cu* atoms for different orientations considered in the present study. It is observed from the figure that for <001> direction, with the increment of alloying element, the dislocation loops are increased up until 5% *Cu*. For 10% *Cu* inclusion in <001> loading direction, the dislocation propagation is less prominent compared to 5% *Cu* inclusion. This decrease of dislocation loop after 5% *Cu* inclusion is also indicated in Fig. 5(a). At 1% *Cu* addition, a separated prismatic loop is visible. The dislocation networks become more dispersed throughout the material with increase in percentage of *Cu*. However, in <110> direction, this kind of extensive dislocation network is not visible, and in some cases, addition of *Cu* can blunt the dislocation loop and hinders its propagation. A comparison between Fig. 6 shows that, the increase in *Cu* weight fraction hinders the propagation of dislocation although the number of dislocations is higher for 2% *Cu* alloy compared to that of 1% *Cu*. At higher percentages of *Cu*, the dislocation loops move towards the bottom of the substrate. Overall, it can be seen from Fig. 6 that new dislocation loops seem to originate rather than lengthening of the loops near the indenter, which is a manifestation of solid solution hardening within the material.[61] From Fig. 6, it is apparent that, dislocation due to inclusion of *Cu* is more prominent in <001> loading direction during nanoindentation.

**3.1.3 Effects of alloying on hardness and reduced modulus**

Variation of hardness and reduced modulus with *Cu* weight percentage are shown for different loading directions, in Fig. 7(a) and (b) respectively. From Fig 7(a), it can be seen that, the variation of hardness with *Cu* percentage follows different trends for different loading directions. For <001> loading direction, the hardness decreases with increasing *Cu* weight percentage. In case of <110> loading direction, 1% *Cu* inclusion results in increased hardness compared to pure *Al*. The hardness decreases with the increase of *Cu* weight percentage after 1% *Cu*. Finally, when the loading direction is <111>, the inclusion of *Cu* up to 2% does not result in decreasing hardness, as opposed to <001> loading direction. Furthermore, 2% *Cu* inclusion in <111> loading direction yields higher hardness compared to pure *Al*. Two key points can be obtained from Fig. 7(a). First one is that, for <111> and <110> loading directions, *Cu* inclusion produces hardening effect in the weight percentage range of 1-2%. And the second one is, inclusion of higher percentages of *Cu* (5% and 10%) results in softening regardless of the loading direction.

Figure 7(b) shows the variation of reduced modulus with weight percentage of *Cu* inclusion for different loading directions. From the figure, it is clear that, inclusion of *Cu* decreases the reduced modulus regardless of the loading direction. Although, this decrease is very prominent in case of <001> and <001> loading directions, compared to <111> direction. Another important observation is that, the reduced modulus for <001> direction is significantly less than that of the other two loading direction cases. These variations of hardness and reduced modulus, with *Cu* percentage and loading direction are very important to understand the feasibility of different applications of *Al-Cu* alloy in nanoscale.

**3.1.4 Effects of alloying on von Mises stress distribution and surface imprint**

In Figs. 8, the von Mises stress distribution after the unloading process are shown for different *Cu* weight fractions and loading directions. For pure *Al*, the stress is mostly localized where the indentation occurred after the unloading process. But in *Al-Cu* alloys, there are presence of a number of localized tensile stress regions that spread across the whole alloy structure. This presence is due to the difference between *Cu* and *Al* atomic sizes, which in turn creates localized tensile stress zones. It is observed that, the stress distribution is non-symmetric. This is due to the fact that, different slip planes become activated and the dislocations are propagated along the different planes disregarding any symmetry. As a result, the stress distribution is different in different positions of the material which clearly indicates the anisotropy of this alloy. The *Cu* atoms present in the alloys (for 1% and 2% *Cu* inclusion), act as anchors for stress and alter the dislocation glide plane and somewhat increase the strength which is reflected in Fig. 7(a). From the unloading results (Fig. 8), it is quite obvious that the residual stress is increasing with the percentage of *Cu* in the alloy. Another observation from Figs. 8 is that, Von Mises stress distribution is more uniformly distributed for alloys with higher percentage of *Cu* (5% and 10%). This indicates that, the material becomes more isotropic with increasing *Cu* inclusion. Hence it can be concluded that the alloying makes the material better at supporting the load throughout the section and not just the plastic deformation within the imprint zone.

The surface imprints for the alloys are shown in the Fig. 9. It is observed that with the higher alloying element percentage, the pile up phenomenon becomes more obvious. This is true for all three directions of <001>, <110> and <111>.

**3.2 Tensile loading of *Al* and *Al-Cu* nanowire**

Tensile test simulations are performed for pure *Al* and *Al-Cu* nanowires made with different percentages of the alloying element to obtain the tensile properties. From the tensile test data, the stress-strain curve is drawn which is used to determine the material properties which are further used for the finite element nanoindentation simulation. In Fig. 10, the stress-strain curves are shown for different percentages of *Cu* and loading in different directions considered in the present study.

For all directions, there is no significant presence of serrations in the stress strain curves after failure, which indicates pure brittle fracture. With the increase of *Cu* percentage, the failure stress and strain are increased for <001> loading direction. As shown in Fig. 10(a), the maximum failure strain and strength are obtained for the 0% *Cu*. For <110> direction, the stress strain diagram shows some minor yielding phenomena before the fracture in case of different *Al-Cu* alloys. Here, the ultimate stress is the highest for pure *Al* with strength decreasing as the amount of *Cu* increases. The fracture strains of *Al-Cu* alloys are also lower than the pure *Al* cases. This result shows that, along <110> direction, alloying *Cu* with *Al* does not necessarily result in improved mechanical properties. For <111> loading direction, we observe yielding before failure similar to <110> direction. In this loading direction, ultimate tensile strength decreases with increasing *Cu* inclusion. It can be summarized that among the three directions, <001> shows the highest materials strength and fracture strain.

The materials properties obtained from these stress-strain curves are presented in Table 2.

**Table 2.** Materials properties of *Al-Cu* alloy at different percentages of *Cu* and at different directions of loading.

| Loading directions | *Cu* (%) | Elastic Modulus (GPa) | Yielding Strain (%) | Yield Stress (GPa) |
|---|---|---|---|---|
| **<001>** | 0 | 53.76 | 8.57 | 5.914 |
| | 1 | 49.60 | 8.07 | 5.605 |
| | 2 | 49.19 | 7.53 | 5.182 |
| | 5 | 49.30 | 6.97 | 4.510 |
| | 10 | 39.29 | 7.01 | 4.118 |
| **<110>** | 0 | 85.00 | 5.50 | 3.067 |
| | 1 | 83.94 | 5.32 | 2.976 |
| | 2 | 82.36 | 5.35 | 2.911 |
| | 5 | 81.61 | 4.66 | 2.601 |
| | 10 | 78.20 | 4.17 | 2.341 |
| **<111>** | 0 | 88.80 | 5.88 | 3.923 |
| | 1 | 87.53 | 5.14 | 3.567 |
| | 2 | 87.48 | 5.21 | 3.597 |
| | 5 | 85.59 | 4.62 | 3.250 |
| | 10 | 80.58 | 4.67 | 3.080 |

## 4. FE results of nanoindentation on *Al-Cu* alloy

### 4.1 Effects of alloying on P-h curves, hardness and reduced modulus

The FE simulations are done considering the nanoindentation problem as a 2D axisymmetric problem for convenience and the potential to solve the problem with less computation effort. The material properties used for FE simulations are obtained from the aforementioned (section 3.2) tensile tests of pure *Al* and *Al-Cu* alloy nanowire using the MD method. Therefore, the same dimension scale has been used in the FE simulations to safeguard from the size effects.[62] Furthermore, the boundary conditions along x and y directions in case of MD simulation are periodic, which means the *Al*-Cu structure used in MD simulations can be considered infinitely extended along x and y directions. Hence, FEM modeling of our present nanoindentation problem is expected to yield reasonable results regardless of nanometer length scale range. *Al* and *Al-Cu* alloy are modeled in FE considering them as isotropic homogeneous materials. From the nanoindentation simulations of the FE analysis, the load displacement curve is obtained and after analyzing the load displacement curve the material properties such as hardness and reduced modulus are calculated. In Fig. 11(a), (b) and (c), the load-displacement curves obtained from the FE simulations for *Al* and *Al−Cu* alloy are shown for indentation in different directions of <001>, <110> and <111>, respectively. It is observed that for <001> direction, the indentation force is maximum for pure *Al* which is similar to the results obtained from the MD simulations. Although the magnitudes of the forces obtained from the FE simulations are very similar to those obtained from MD simulations, the trends obtained from both the methods are different. This could be due to the difference in boundary conditions in the two approaches. Although, the boundaries are assumed to be periodic in MD, that is not the case in FEM. The discrepancies may also be for assuming the material is axisymmetric.

From the load-displacement curve obtained from the FE simulations, the material properties such as hardness and reduced modulus have been extracted following the same method used in the MD part of the study. The variation of hardness and reduced modulus with *Cu* percentage is shown in Fig. 12 for different loading directions. It can be observed from the Fig. 12(a) that, the values of hardness, obtained from both the methods, are very similar in magnitudes for all the three loading directions and different *Cu* percentages. Also, the variation of hardness obtained from FE simulations follow similar trend obtained from MD simulations. On the other hand, Fig. 12(b) shows that, the variation trend of reduced modulus with *Cu* percentage extracted from FE simulations is similar to that obtained from MD simulations. However, the values of reduced modulus are significantly different from MD results.

**4.2 Comparison of FE results with MD results**

The primary motivation of performing the FE analysis is to obtain the material properties with less computational time and cost. To validate the approach, results obtained by the FE simulations are compared with the previous results of MD.

The comparison of MD and FE load-displacement curves for the *Al−Cu* alloy consisting 1% *Cu* in <001>, <110> and <111> loading directions are presented in Fig. 13. The load-displacement curves show good agreement between MD and FE results. It is observed from these figures that there are some fluctuations in the MD indentation curves. These fluctuations are due to the plastic event such as dislocation nucleation, interactions and propagation inside the materials. However, the curves obtained from the FE simulations do not show such variation as the material is modeled considering an isotropic material. In MD, the material yielding, and plastic phenomenon can be more accurately predicted than the FE modeling. However, considering the fact that MD takes longer computation time and is limited in length scale, applying FE analysis for nanoindentation problems are useful, especially when just considering the material properties such as hardness and elastic modulus and not the dislocation mechanisms. For <110> direction, FE can successfully predict the load-displacement relation well up to indentation depth of 0.2 nm displacement in the $P-h$ curves. In case of <111> direction, FE P-h curve conforms with MD up to 1 nm. In Fig. 13 the slopes of the unloading curves for <110> and <111> loading directions are also similar to the MD results. As a result, the obtained material properties are very close and in good agreement with each other.

The results obtained by the MD and FEM are shown for *Al−Cu* in Table 3 showing that the obtained hardness results are in good agreement.

**Table 3.** Comparison of hardness for *Al-Cu* alloy at different percentage of *Cu* and at different direction of loading obtained by MD and FEM simulations.

| Direction | *Cu* Percentage | Hardness (MD) (GPa) | Hardness (FEM) (GPa) | Percentage Deviation (%) |
|---|---|---|---|---|
| <001> | 0 | 7.817 | 7.760 | 0.73 |
| | 1 | 7.757 | 7.236 | 6.72 |
| | 2 | 7.511 | 6.986 | 6.99 |
| | 5 | 6.921 | 6.618 | 4.38 |
| | 10 | 6.498 | 5.969 | 8.14 |
| <110> | 0 | 7.583 | 6.840 | 9.80 |
| | 1 | 7.641 | 6.672 | 12.68 |
| | 2 | 7.574 | 6.530 | 13.78 |
| | 5 | 7.256 | 6.222 | 14.25 |
| | 10 | 7.373 | 5.946 | 19.35 |
| <111> | 0 | 7.564 | 8.187 | 8.24 |
| | 1 | 7.532 | 7.669 | 1.82 |
| | 2 | 7.571 | 7.710 | 1.84 |
| | 5 | 7.132 | 7.144 | 0.17 |
| | 10 | 7.177 | 6.758 | 5.84 |

## 5. Conclusions

In the present study, nanoindentation simulations are performed for pure *Al* and *Al-Cu* alloys of different *Cu* weight percentages and in different loading directions. Both FE simulations and MD simulations are carried out to calculate the material properties such as hardness and reduced modulus. The input parameters of the FE simulations, have been obtained by performing tensile tests on pure *Al* and *Al-Cu* nanowires using MD simulations. The FE nanoindentation simulations are then performed, and their results have been compared with the MD simulation results where good agreement has been found in case of <110> and <111> loading directions. The technique of implementing finite element modelling at nanoscale can be a low fidelity alternate of MD simulations because significantly less computational time is required. From the results, the following conclusions can be drawn:

- For *Al-Cu* alloy, the addition of *Cu* (up to 2%) results in solid solution strengthening and creates local strain field inside the *Al*. This strain field creates energy barrier for the dislocations to move along its slip plane and forces the dislocations to choose energetically favorable slip plane at a higher strength than previous. The hardness and reduced modulus values vary with the different percentage of *Cu* for different crystallographic directions.

- Addition of *Cu* with higher weight percentages (5% and 10%) results in softening of the *Al-Cu* alloy.
- We have observed that there are no dislocations during the initial indentation period, in case of all loading direction for any of the *Al-Cu* alloys.
- For <001> and <111> loading direction, the dislocation loops increase with increment of alloying element, and also become more dispersed throughout the material. As a result, dislocation densities and hence solid solution strengthening effects are maximum for these loading directions.
- In *Al-Cu* alloys, the stress distribution is different in different positions of the material, since the distribution of *Cu* atoms in the structure is random, and the *Cu* atoms act as the stress anchors which makes the alloy anisotropic.
- Implementing the MD tensile test data makes it possible to obtain the hardness and reduced modulus value from FEM and compare it with the MD results. The comparison showed that the load-displacement curves obtained for all the *Al-Cu* alloys are in fairly good agreement with the MD results for all loading directions, considering that FE simulations is a macroscopic view into this nanoscale phenomena. In this way, our present method can be a good low-fidelity alternative for high-fidelity MD simulations.

**Acknowledgement**

The authors acknowledge the high-performance computing facilities provided by the IICT, BUET during this study. The authors also thankful to the department of Mechanical Engineering, BUET for their support. The authors are grateful to Professor Dibakar Datta, Professor Md. Mahbubul Islam, and Professor H. M. Mamun *Al* Rashed for many useful discussions on the results.

**List of Figure Captions**

| | |
|---|---|
| **Figure 1** | Physical modeling of the nanoindentation problem. The indenter is analytically rigid as modeled in LAMMPS |
| **Figure 2** | Figure 2. Modeling of the alloy in different percentage of *Cu*. |
| **Figure 3** | Schematic of the axisymmetric modeling of the nanoindentation in the FEM. |
| **Figure 4** | Load displacement curve for different *Cu* percentage during indentation in (a) <001>, (b) <110>, (c) <111> direction of *Al−Cu* alloy. |
| **Figure 5** | Variation in dislocation density for different *Cu* percentage during indentation in (a) <001>, (b) <110>, (c) <111> direction of *Al−Cu* alloy |
| **Figure 6** | Dislocation nucleation for different *Cu* percentage after maximum loading in <001>, <110> and <111> direction of *Al−Cu* alloy |
| **Figure 7** | Variation of (a) hardness and (b) elastic modulus for different *Cu* percentage in *Al−Cu* alloy and loading direction |
| **Figure 8** | Von Mises stress distribution for different *Cu* percentage after unloading in <001>, <110> and <111> direction of *Al−Cu* alloy. |
| **Figure 9** | Surface imprint for different *Cu* percentage after maximum loading in <001>, <110> and <111> direction of *Al−Cu* alloy. |
| **Figure 10** | Stress-strain relationship for different *Cu* percentage during tensile loading in (a) <001>, (b) <110>, (c) <111> direction of *Al-Cu* alloy. |
| **Figure 11`** | Load-displacement curve in (a) <001>, (b) <110>, (c) <111> direction of Al-*Cu* alloy. |
| **Figure 12** | Variation of (a) hardness and (b) elastic modulus for different *Cu* percentage in Al-*Cu* alloy and loading direction. |
| **Figure 13** | Comparison of FEM and MD load displacement curve for loading in (a) <001>, (b) <110>, (c) <111> direction of 1% *Al-Cu* alloy. |

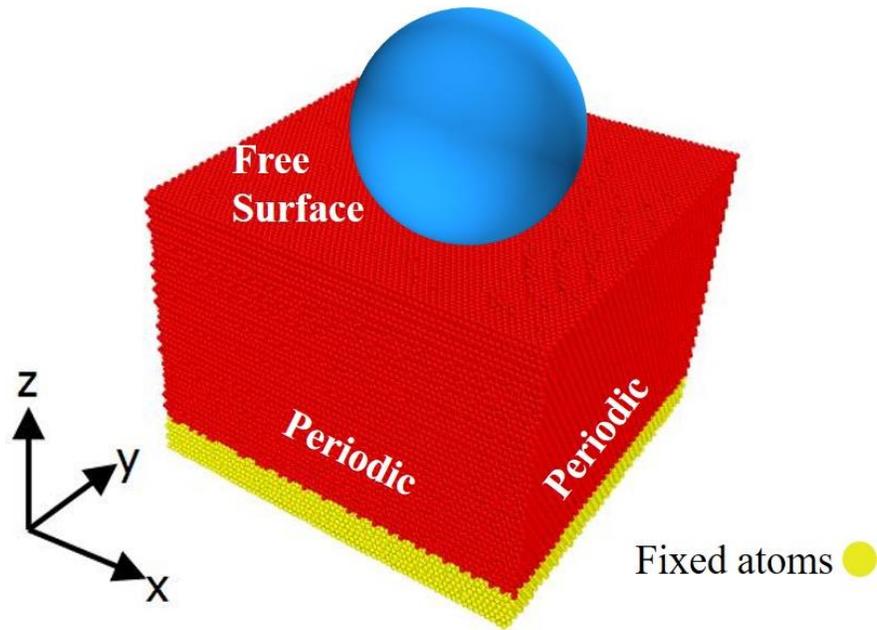

Figure 1. Physical modeling of the nanoindentation problem. The indenter is analytically rigid as modeled in LAMMPS

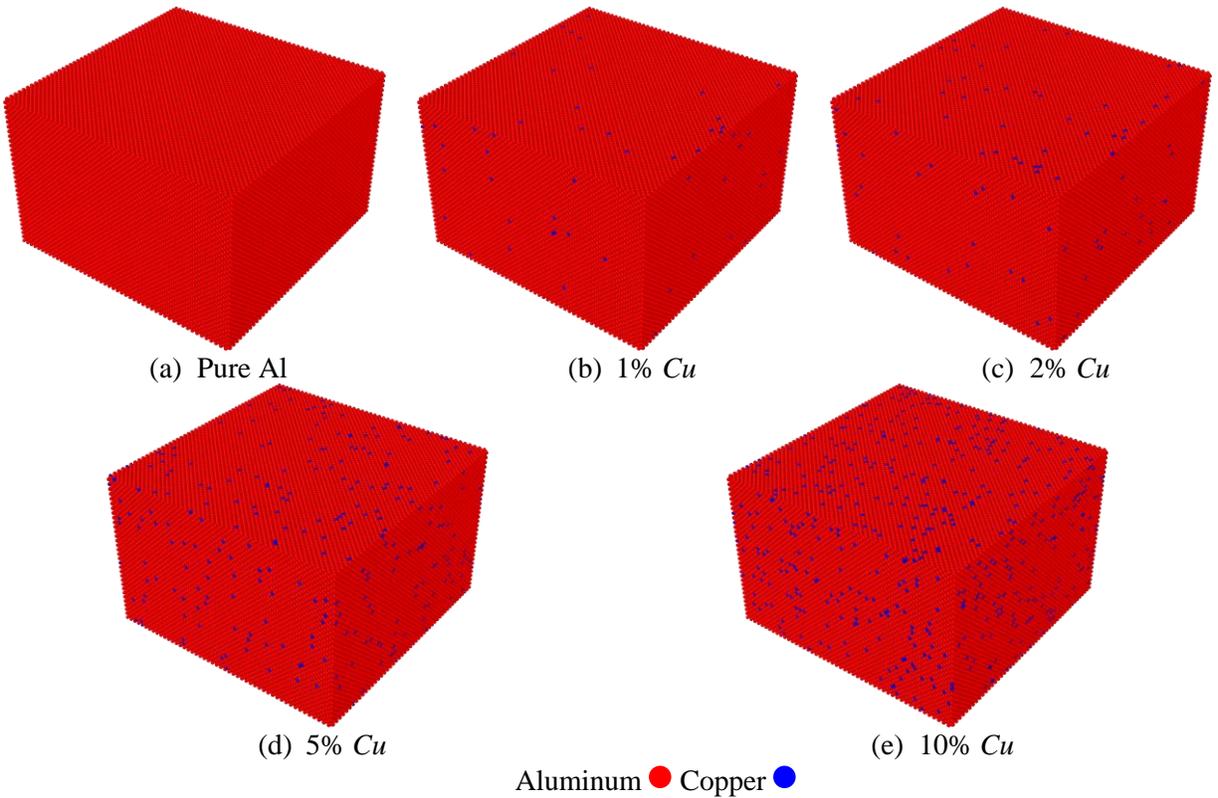

Figure 2. Modeling of the alloy in different percentage of *Cu*.

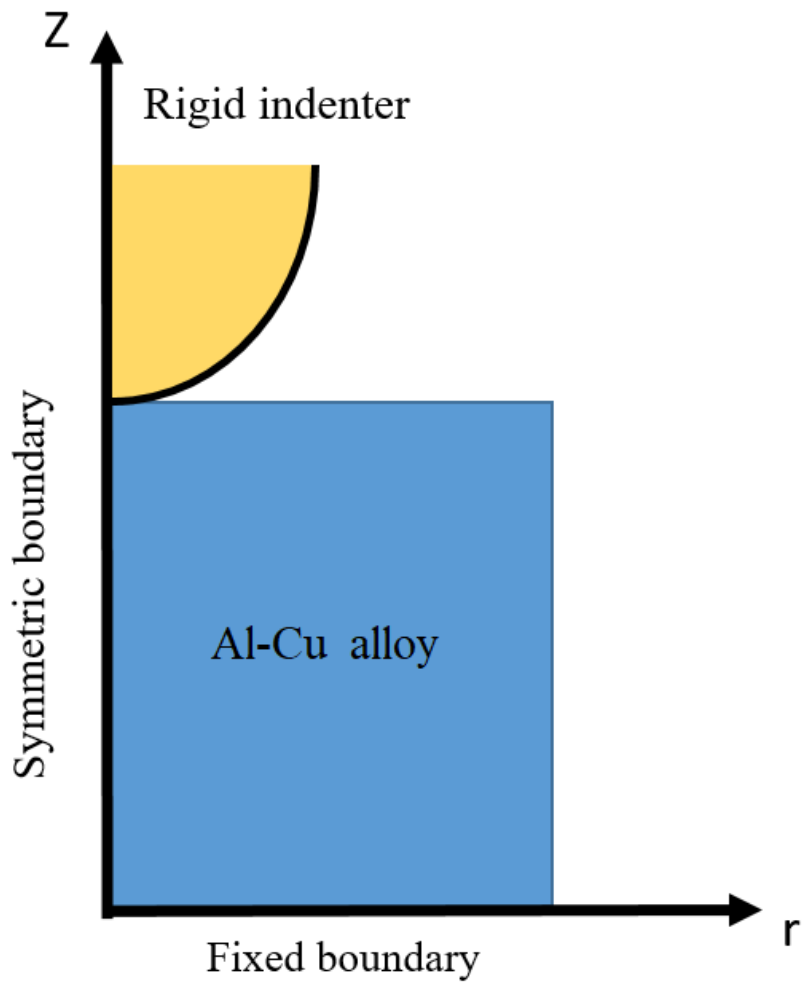

Figure 3. Schematic of the axisymmetric modeling of the nanoindentation in the FEM.

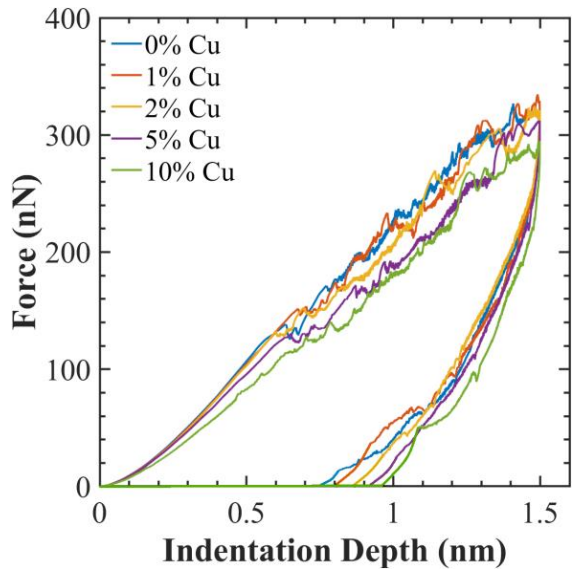
(a)

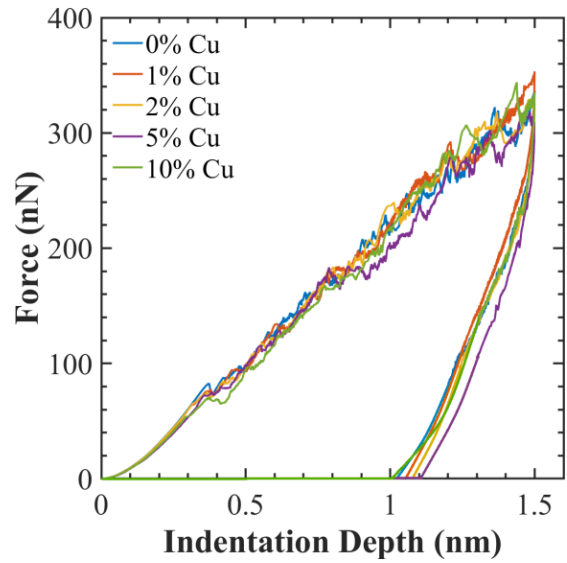
(b)

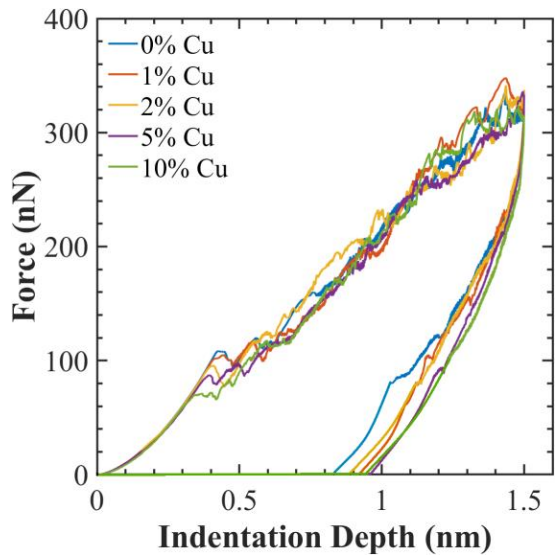
(c)

Figure 4. Load displacement curve for different *Cu* percentage during indentation in (a) <001>, (b) <110>, (c) <111> direction of *Al−Cu* alloy.

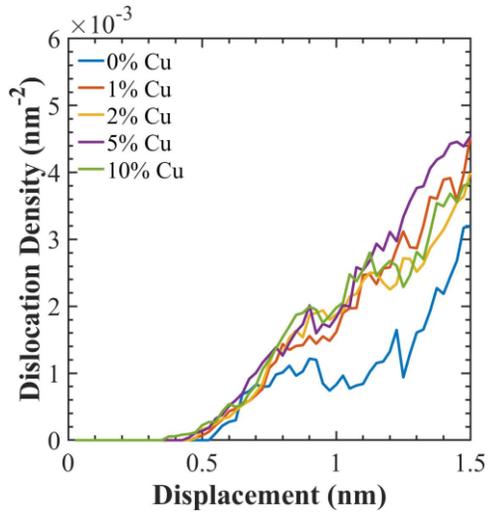
(a)

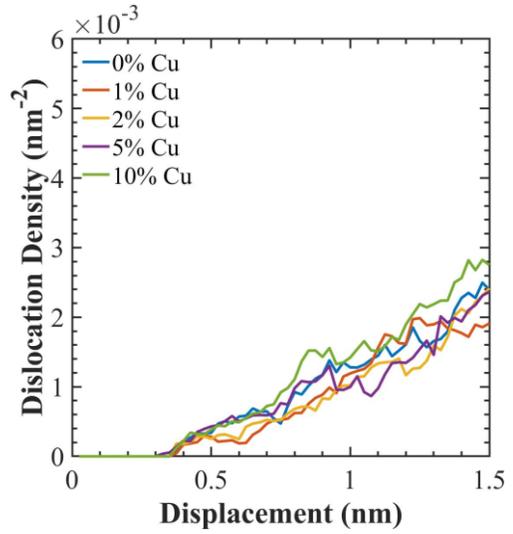
(b)

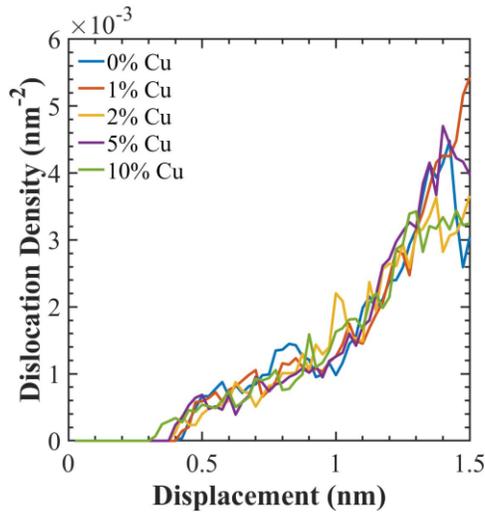
(c)

Figure 5. Variation in dislocation density for different *Cu* percentage during indentation in (a) <001>, (b) <110>,

(c) <111> direction of Al−*Cu* alloy

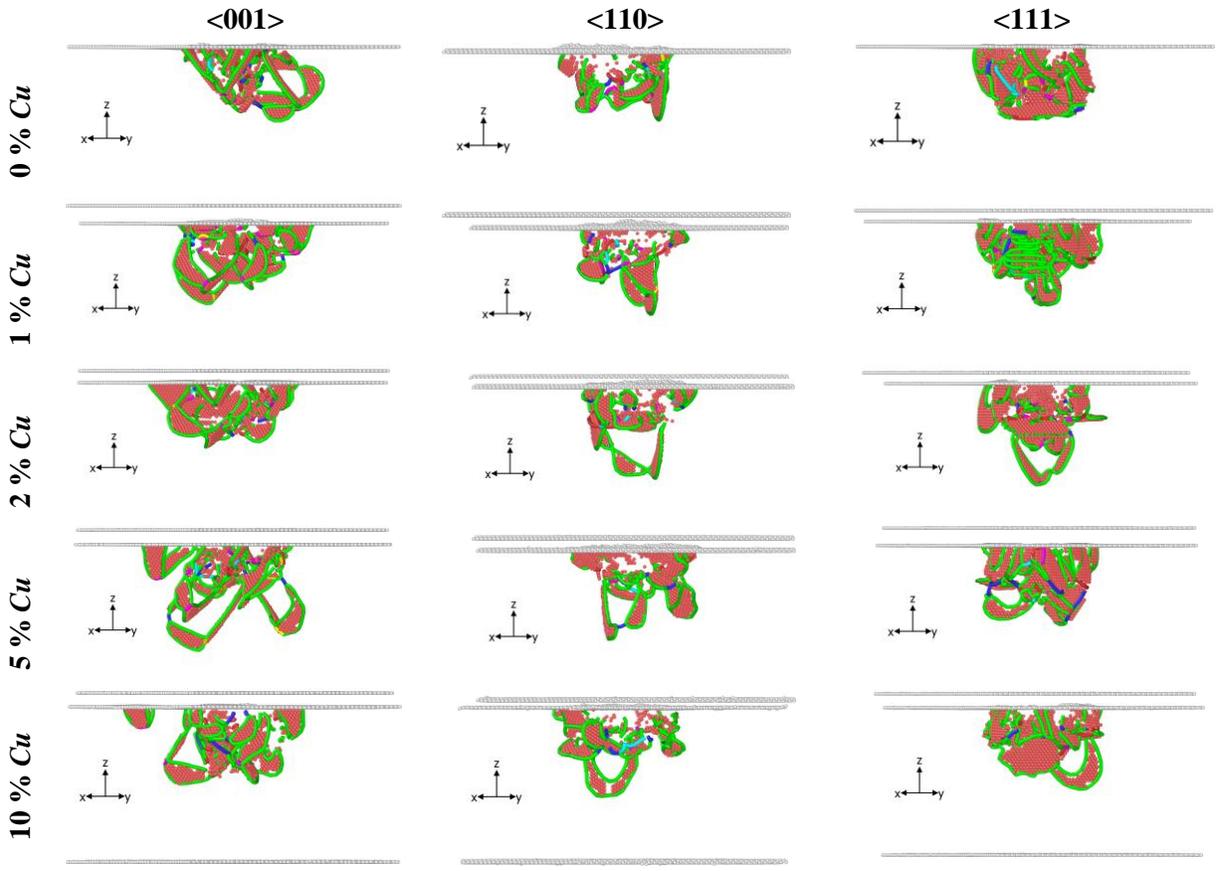

Figure 6. Dislocation nucleation for different *Cu* percentage after maximum loading (1.5nm depth) in <001>, <110> and <111> direction of *Al−Cu* alloy

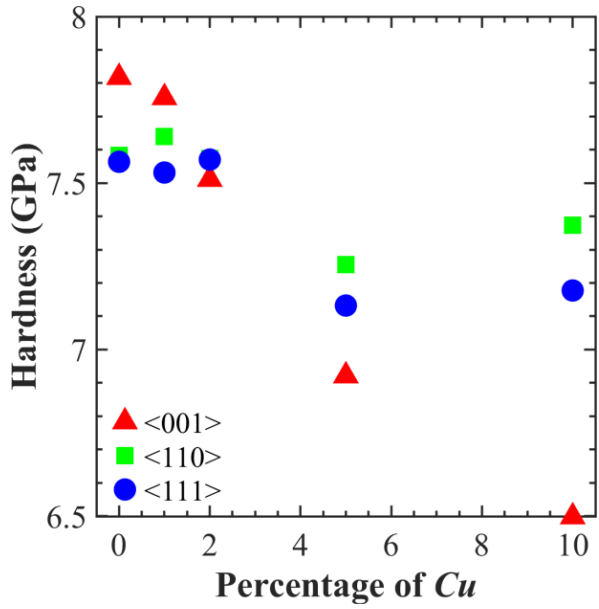 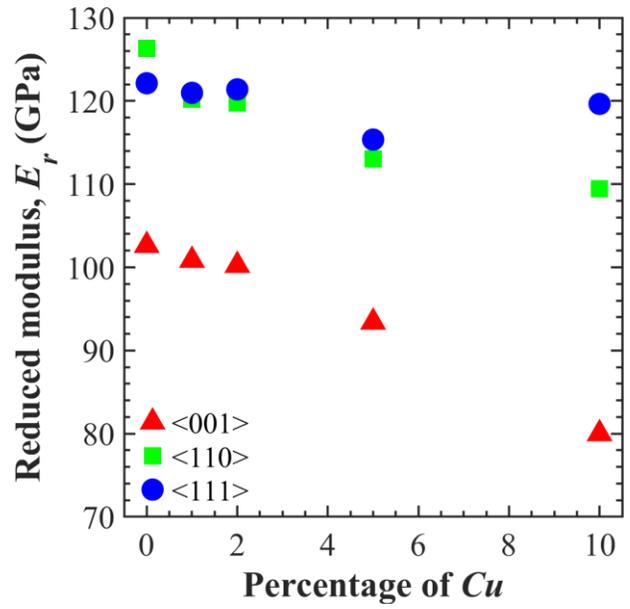

(a)                  (b)

Figure 7. Variation of (a) mean hardness and (b) reduced modulus for different *Cu* percentage in *Al−Cu* alloy and loading direction.

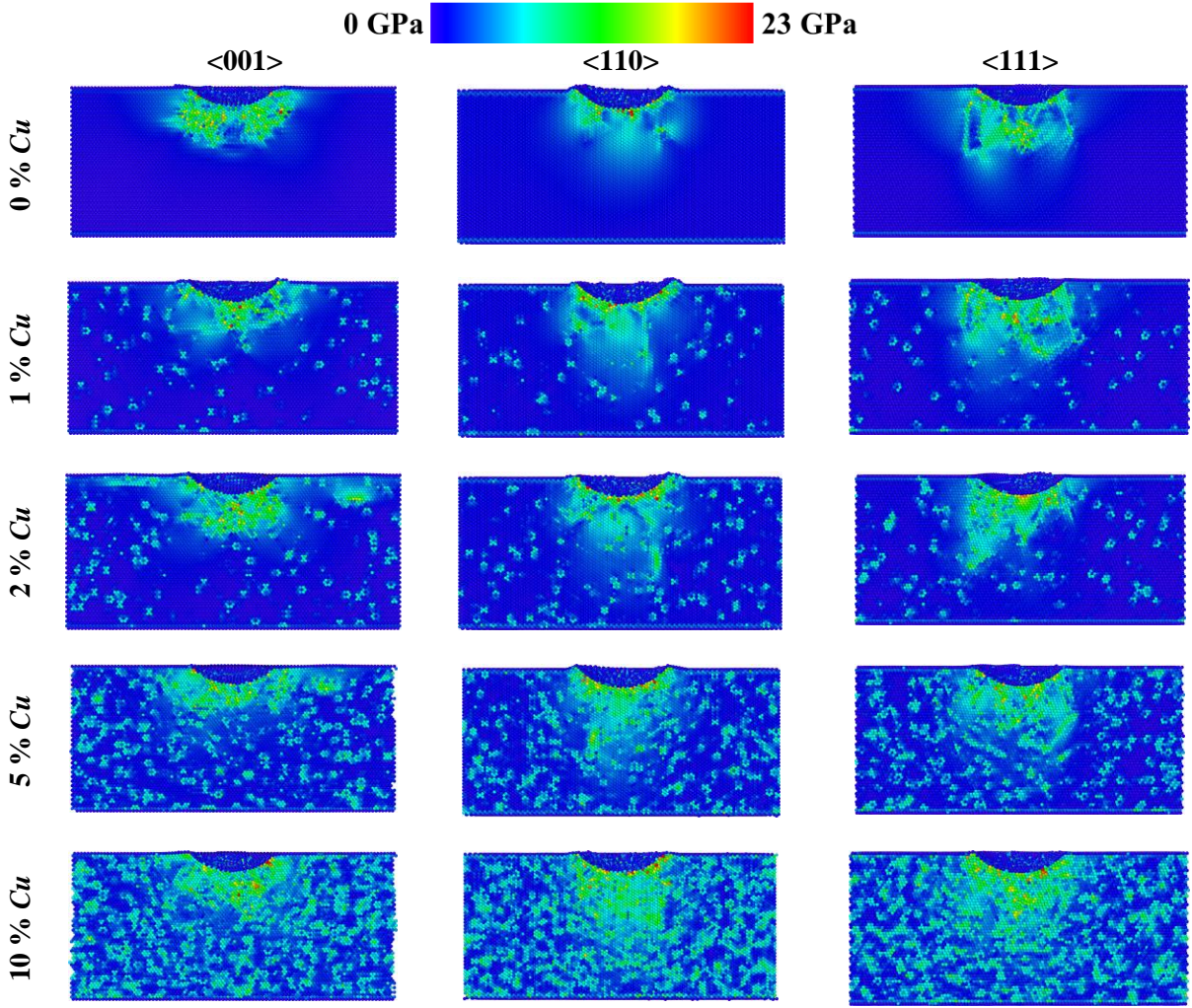

Figure 8. Von Mises stress distribution for different *Cu* percentage after unloading in <001>, <110> and <111> direction of *Al−Cu* alloy

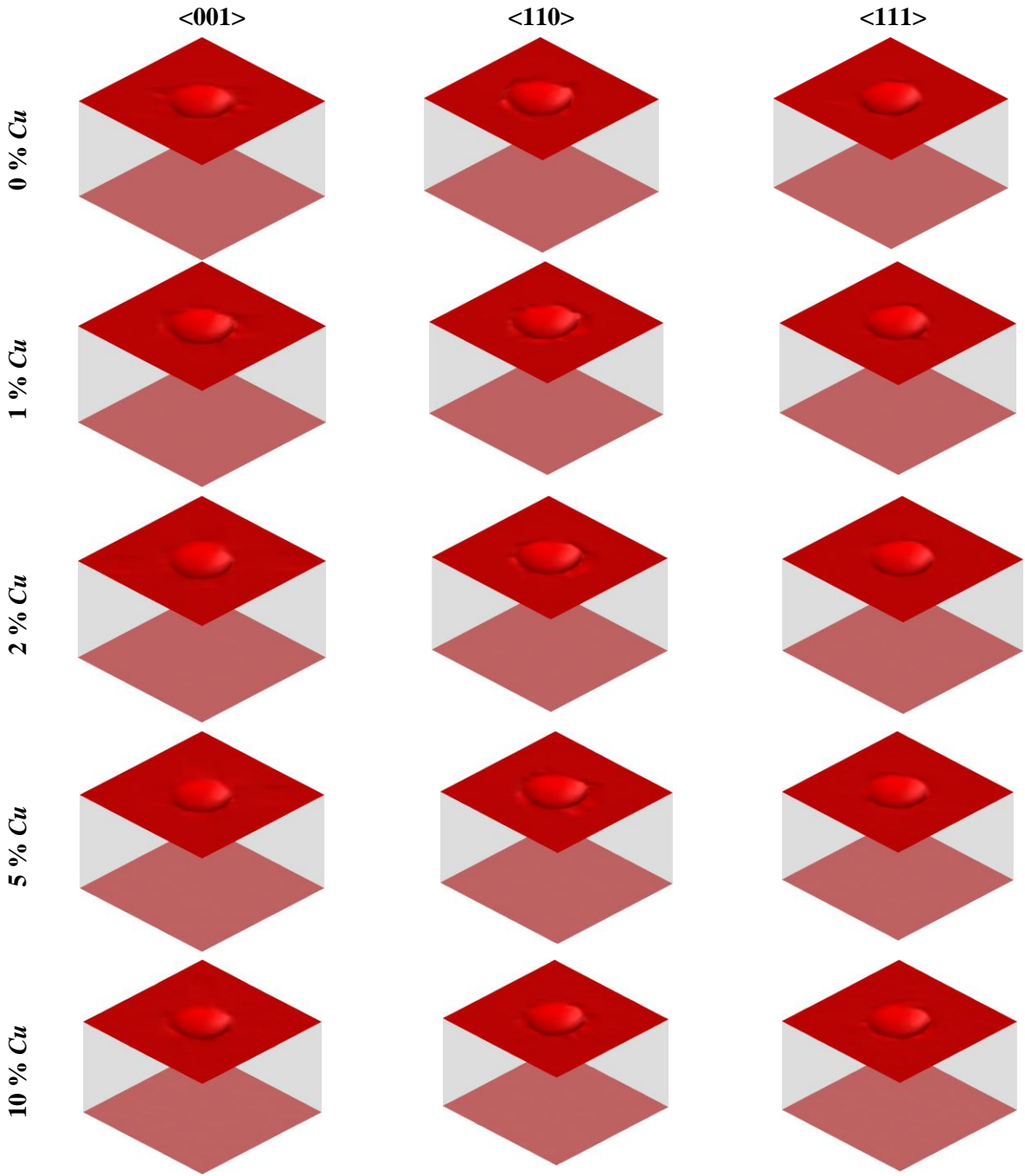

Figure 9. Surface imprint for different *Cu* percentage after maximum loading (1.5nm depth) in <001>, <110>, and <111> direction of *Al−Cu* alloy.

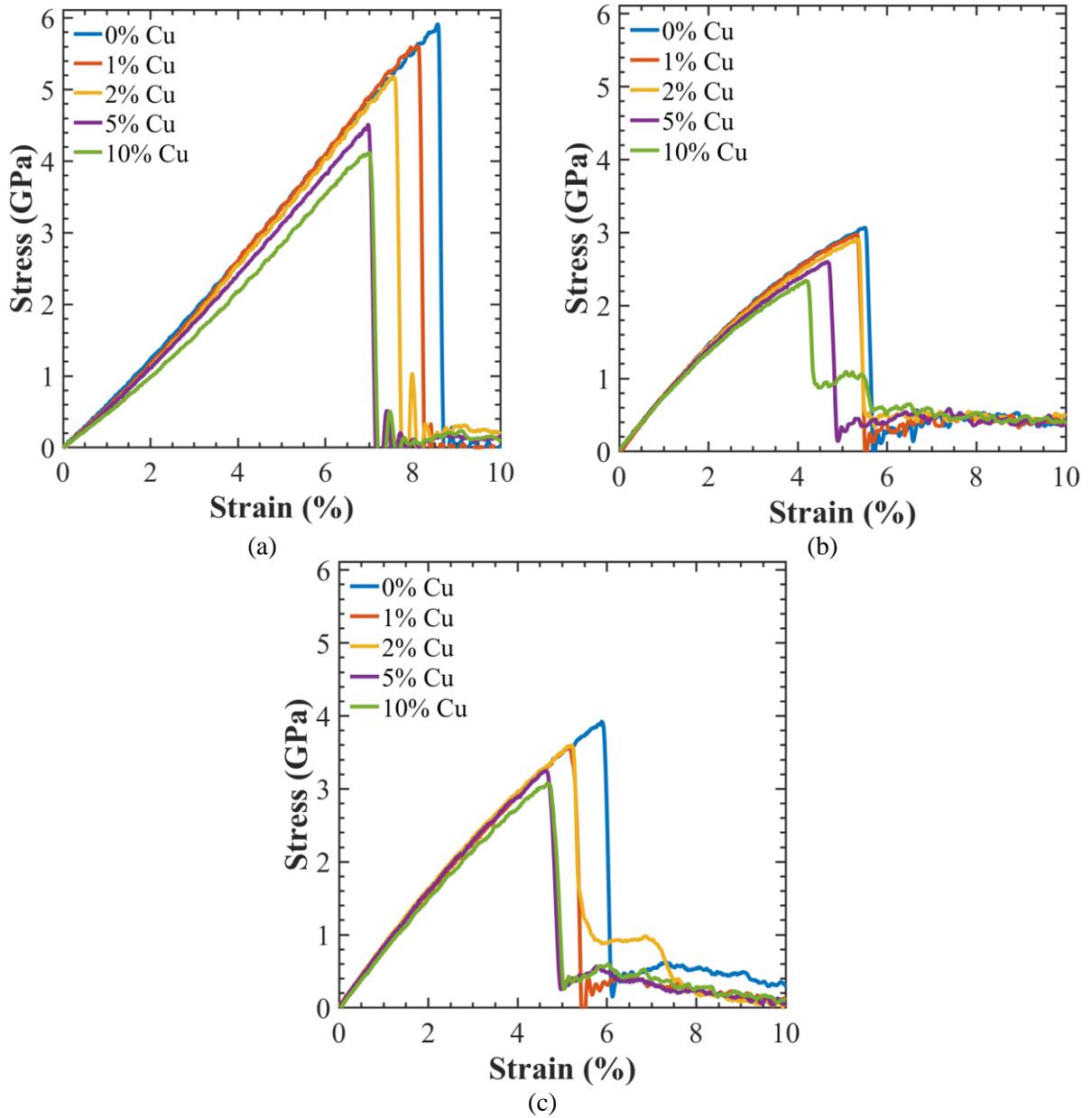

Figure 10. Stress-strain relationship for different *Cu* percentage during tensile loading in (a) <001>, (b) <110>, (c) <111> direction of *Al − Cu* alloy.

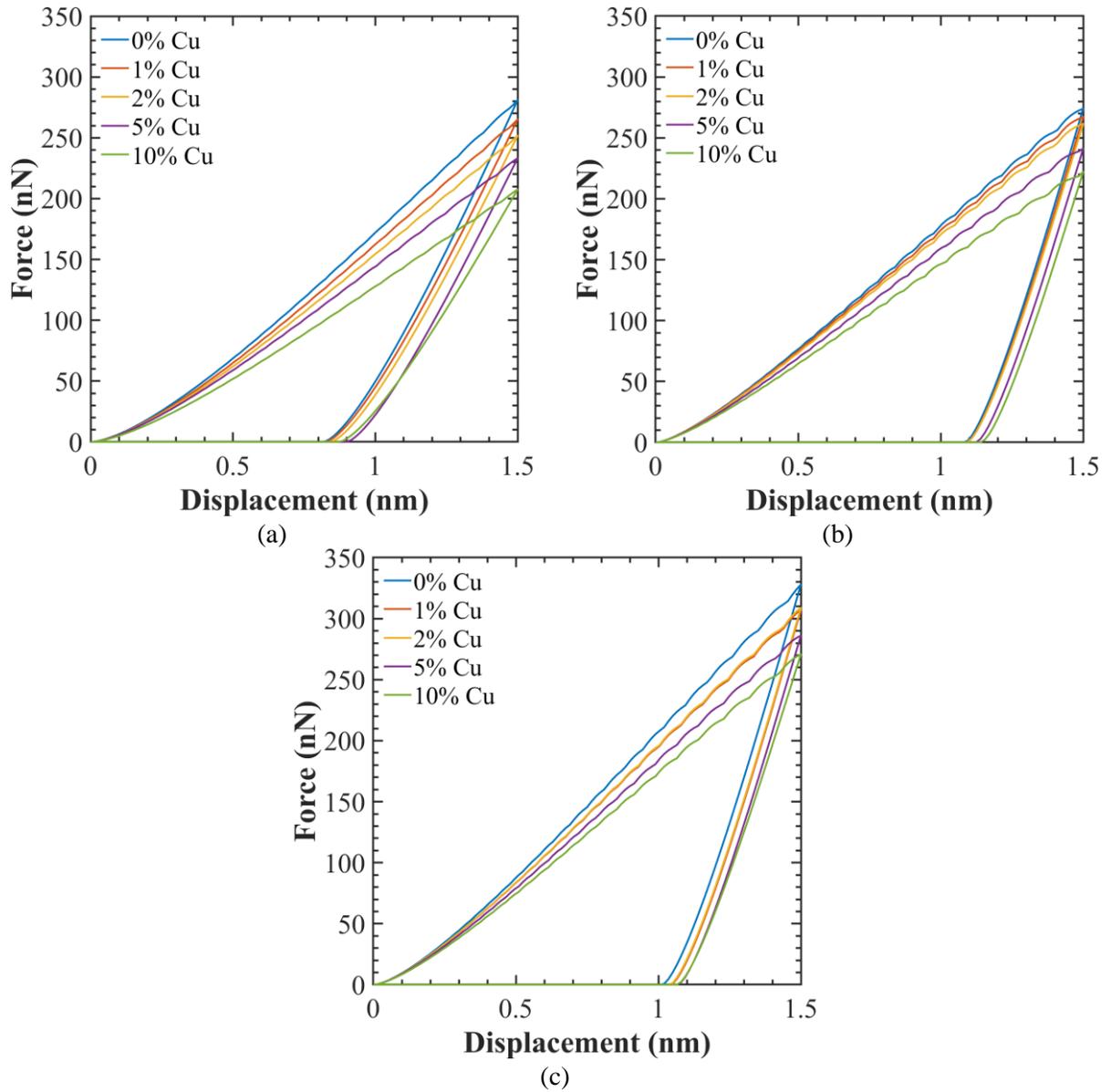

Figure 11. Load-displacement curve in (a) <001>, (b) <110>, (c) <111> direction of Al-*Cu* alloy using eam/fs potential.

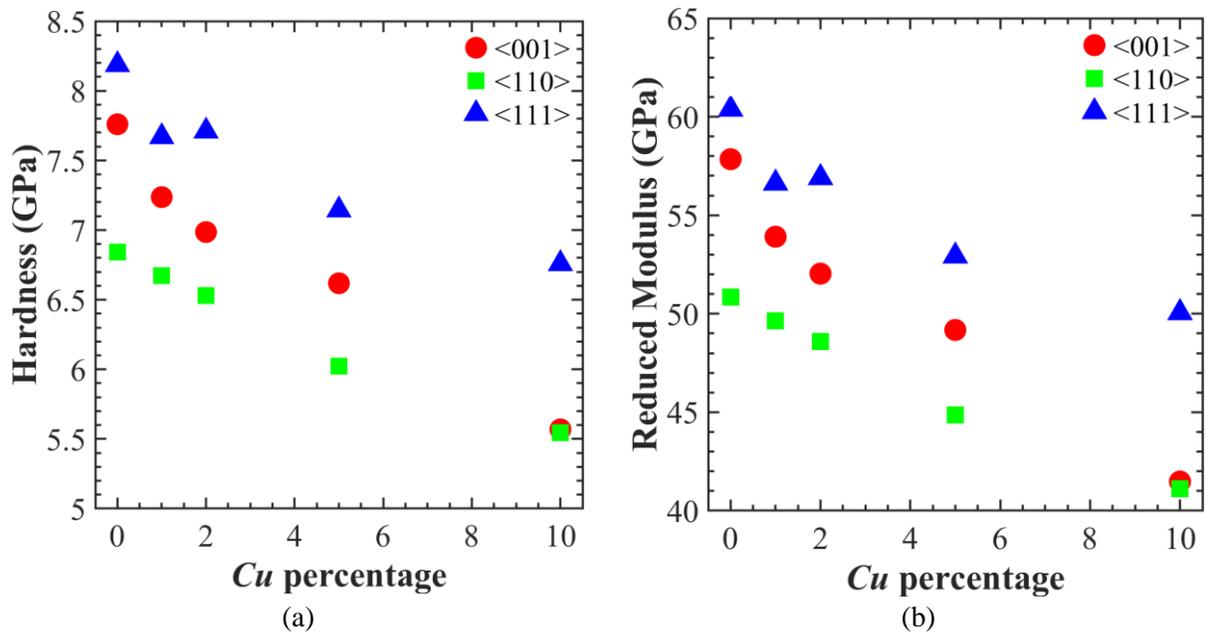

Figure 12. Variation of (a) hardness and (b) elastic modulus, as obtained from the FE simulations, for different *Cu* percentage in *Al-Cu* alloy and loading direction.

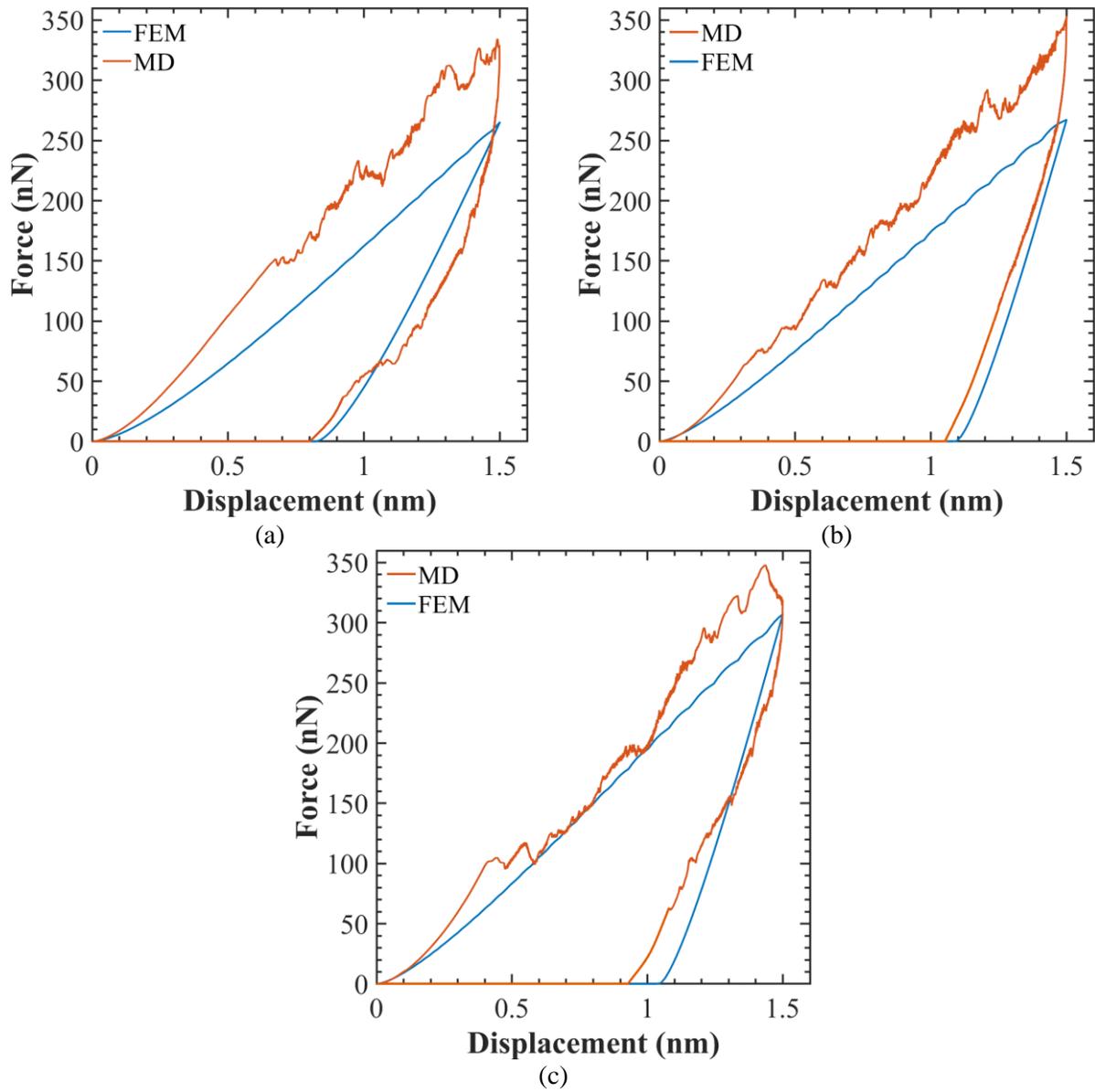

Figure 13. Comparison of FEM and MD load displacement curve for loading in (a) <001>, (b) <110>, (c) <111> direction of 1% *Al-Cu* alloy.